\documentstyle[preprint,prb,aps]{revtex}
\def\bdown{\begin{array}{c}-\\[-3mm]\downarrow\end{array}}

\def\tup{\begin{array}{c}\uparrow\\[-3mm] -\end{array}}
\def\tdown{\begin{array}{c}\downarrow\\[-3mm]-\end{array}}
\def\bup{\begin{array}{c}-\\[-3mm]\uparrow\end{array}}
\def\bdown{\begin{array}{c}-\\[-3mm]\downarrow\end{array}}

\begin{document}

\title{Low temperature properties of one-dimensional
SU(4) Hubbard-like model at low concentration}
\author{Shi-Jian Gu}
\address{
Zhejiang Institute of Modern Physics,
Zhejiang University, Hangzhou,310027, China
}

\author{You-Quan Li}
\address{
Zhejiang Institute of Modern Physics,
Zhejiang University, Hangzhou,310027, China\\
Institute for Physics, Augsburg University, D-86135 Augsburg, Germany
}
\author{Zu-Jian Ying}
\address{
Zhejiang Institute of Modern Physics,
Zhejiang University, Hangzhou,310027, China
}

\date{Received: }
\maketitle

\begin{abstract}
On the basis of Bethe ansatz solution of one dimensional SU(4)
Hubbard-like model, we study its thermodynamics properties by means
of Yang-Yang thermodynamics Bethe ansatz.
The Land\'e $g$ factor is taken into account so as to describe electrons
with orbital degeneracy.
The free energy at low temperature is
given and the specific heat both in strong coupling and
weak coupling limits are obtained.
\end{abstract}

\section{Introduction}

The study of integrable model has a long and rich history in 
condensed matter physics beginning with Bethe's solution
of the one-dimensional Heisenberg Model and extending
to nowadays when a variety of soluble models provide the paradigms 
that enrich much of our physical intuition. 
Recently there has been much interests in the studies on the $3d$ 
electrons in transitional metal oxide \cite{TNagaosa}
because of the existence of orbital degree of freedom\cite{McWhan,Pen} 
in addition to the spin one.
Since the SU(4) symmetry was pointed out \cite{LiMSZ98}, there has 
been various studies, such as 
critical properties in photoemission spectra for the one-dimensional 
Mott insulator with orbital degeneracy \cite{FTKawawami}.
One dimensional solutions of the model in insulating limit 
\cite{LiMSZ99} as well as the Hubbard-like model \cite{LiGYE00}
were presented. 
There were discussions of the thermodynamics based on the SU(N) 
generalized Bethe ansatz equation \cite{FrahmSS},
but the physical meaning of the external field is ambiguous there
because it is not associated with spin and orbital directly.

In present paper, we study the low-temperature thermodynamics of SU(4) 
Hubbard-like model for electrons 
with 2-fold orbital degeneracy at low concentration. 
The thermal equilibrium is discussed exactly on the basis of 
the known Bethe-ansatz equations by taking account of the Land\'e $g$ factor. 
Next section we first briefly exhibit the
Bethe-ansatz equation under consideration, then discuss the
thermodynamics of the model by considering string hypothesis.
In Sec. \ref{sec:3}, we calculate the formal expressions of the
thermodynamics quantities, such as free energy etc.. 
In Sec. \ref{sec:4}, the case of low temperature is discussed extensively.

\section{The model and its spectrum in thermodynamics limit}

The model Hamiltonian reads
\begin{equation}
{\cal H}=-t\sum_{i,a}{\cal P}(C_{i,a}^+C_{i+1,a}+C_{i+1,a}^+C_{i,a}){\cal P}
+U\sum_{i,a<a^\prime}
 n_{i,a}n_{i,a^\prime}
 \label{Hamiltonian}
\end{equation}
where $i=1,2,\dots,L$ identify the lattice site, 
and $a=1,2,...,4$ labels the four states of spin and orbital \cite{LiMSZ98}. 
The ${\cal P}$ projects the Hilbert space
 onto the sector that the sites are only occupied by at
most two electrons.
The internal degree of freedom (\ref{Hamiltonian}) 
is specified to spin and orbital in present model.
The Bethe-ansatz equation for the spectrum was 
suggested \cite{Choy,LiGYE00} as
\begin{eqnarray}
  e^{ik_j L} &=& \prod_{b=1}^M\frac{\sin k_j-\lambda_b+i\eta}
   {\sin k_j-\lambda_b-i\eta},\nonumber \\
  \prod_{l=1}^N\frac{\lambda_a-\sin k_l+i\eta}{\lambda_a-\sin k_l-i\eta}
    &=&
-\prod_{b=1}^M\frac{\lambda_a-\lambda_b+i2\eta}{\lambda_a-\lambda_b-i2\eta}  
   \prod_{c=1}^{M'}\frac{\mu_c-\lambda_a+i\eta}{\mu_c-\lambda_a-i\eta},
      \nonumber \\  
\prod_{b=1}^M\frac{\mu_a-\lambda_b+i\eta}{\mu_a-\lambda_b-i\eta}
 &=&-\prod_{c=1}^{M'}\frac{\mu_a-\mu_c+i2\eta}{\mu_a-\mu_c-i2\eta}     
  \prod_{d=1}^{M''}\frac{\nu_d-\mu_a+i\eta}{\nu_d-\mu_a-i\eta}, 
   \nonumber\\   
\prod_{b=1}^{M'}\frac{\nu_a-\mu_b+i\eta}{\nu_a-\mu_b-i\eta}
    &=& -\prod_{c=1}^{M''}\frac{\nu_a-\mu_c+i2\eta}{\nu_a-\mu_c-i2\eta}.
\label{BAE}
\end{eqnarray}
where $\eta=U/4t$. Unlike the conventional SU(2) Hubbard model
the SU(4) generalization (\ref{BAE}) of 
Lieb-Wu solution \cite{LiebW} is valid at low temperature and 
low concentration \cite{ChoyHaldane,FrahmSS}.

For ground state (i.e., at zero temperature), 
the $k,\lambda, \mu,\nu$ are real 
roots of the Bethe ansatz equation (\ref{BAE}). 
However, for the excited state, they can be complex roots.
We will not take account of the complex roots in the charge space $k$ 
for repulsive interaction\cite{Takahashi}.
The complex root of $\lambda,\mu,\nu$ are always form a ``bound state" 
with several same kind of  rapidity which arise from the consistency 
of both hand side of the Bethe ansatz equations.
The complex roots are defined as 
\begin{eqnarray}
\Lambda_a^{nj} &=& \lambda_a^n+(n+1-2j)i\eta, \;\;\;j=1,2\dots,n,\nonumber\\
U_b^{mj} &=& \mu_b^m+(m+1-2j)i\eta, \;\;\;j=1,2\dots,m,\nonumber\\
V_c^{lj}&=& \nu_c^l+(l+1-2j)i\eta, \;\;\;\;\;\;
j=1,2\dots,l,\label{StringConjecture}
\end{eqnarray}
up to the order $O(e^{-L})$ which vanishes in the thermodynamic limit.
Substituting those strings into Eq. (\ref{BAE}) and taking the logarithm
of it, we get
\begin{eqnarray}
2\pi I_j &=& k_j L+2\sum_{an}\tan^{-1}\Bigl(\frac{\sin k_j-\lambda_a^n}{n\eta}\Bigr),
  \nonumber \\
2\pi J_a^n &=& 2\sum_l\tan^{-1}\Bigl(\frac{\lambda_a^n-\sin k_l}{n\eta}\Bigr)
 -2\sum_{bml}A_{nml}\tan^{-1}\Bigl(\frac{\lambda_a^n-\lambda_b^m}{l\eta}\Bigr)
  -2\sum_{clt}B_{nlt}\tan^{-1}\Bigl(\frac{\mu_c^l-\lambda_a^n}{t\eta}\Bigr),
   \nonumber \\
2\pi K_a^n &=& 2\sum_{blt}B_{nlt}\tan^{-1}\Bigl(\frac{\mu_a^n-\lambda_b^l}{t\eta}\Bigr)
  -2\sum_{clt}A_{nlt}\tan^{-1}\Bigl(\frac{\mu_a^n-\mu_c^l}{t\eta}\Bigr)
   -2\sum_{dlt}B_{nlt}\tan^{-1}\Bigl(\frac{\nu_d^l-\mu_a^n}{t\eta}\Bigr),
    \nonumber \\
2\pi Q_a^n &=& 2\sum_{blt}B_{nlt}\tan^{-1}\Bigl(\frac{\nu_a^n-\mu_b^l}{t\eta}\Bigr)
  -2\sum_{clt}A_{nlt}\tan^{-1}\Bigl(\frac{\nu_a^n-\nu_c^l}{t\eta}\Bigr).
    \label{SeqEquations}
\end{eqnarray}
where
\begin{eqnarray}
A_{nmp}&=&\left\{
  \begin{array}{ll}
     1, & {\rm for}\; p=m+n, |m-n|\,(\neq 0),\\
     2, & {\rm for}\; p=n+m-2, n+m-4,\cdots,|n-m|+2,\\
     0, & {\rm  otherwise,}
  \end{array}\right.
   \nonumber\\
B_{nlt}&=&\left\{
  \begin{array}{ll}
    1, & {\rm for}\; t=n+l-1, n+l-3,...,|n-l|+1,\\
    0, & {\rm  otherwise.}
   \end{array}\right.\nonumber
\end{eqnarray}
Noe we consider the question in  the thermodynamic limit
$N,L,M,M',M''\rightarrow \infty$ with a fixed concentration $D=N/L$
by introducing the distribution of roots and holes for $k, \lambda, \mu, \nu$
respectively:
\begin{eqnarray}
\frac{1}{L}\frac{dI(k)}{dk}=&&\rho(k)+\rho^h(k),\nonumber \\
 \nonumber
\frac{1}{L}\frac{dJ_n(\lambda)}{d\lambda}=&&\sigma_n(\lambda)
+\sigma_n^h(\lambda),\\
 \nonumber
\frac{1}{L}\frac{dK_n(\mu)}{d\mu}=&&\omega_n(\mu)+\omega_n^h(\mu),\\
\frac{1}{L}\frac{dQ_n(\nu)}{d\nu}=&&\tau_n(\nu)+\tau_n^h(\nu).
 \label{DOR}
\end{eqnarray}
We obtain the following coupled integral equations.
\begin{eqnarray}
\rho+\rho^h=&& \frac{1}{2\pi}
 +\sum_n\cos k\int K_n(\sin k-\lambda)\sigma_n(\lambda) d\lambda,
  \nonumber \\
\sigma_n+\sigma_n^h=&&\int K_n(\lambda-\sin k)\rho(k)dk
 -\sum_{ml}A_{nml}\int K_l(\lambda-\lambda^\prime)
  \sigma_m(\lambda^\prime)d\lambda^\prime
   +\sum_{lt}B_{nlt}\int K_t(\lambda-\mu)\omega_l(\mu)d\mu,
   \nonumber\\
\omega_n+\omega_n^h=&&\sum_{lt}B_{nlt}\int K_t(\mu-\lambda)
 \sigma_l(\lambda)d\lambda
  -\sum_{lt}A_{nlt}\int K_t(\mu-\mu^\prime)\omega_l(\mu^\prime)d\mu'
  +\sum_{lt}B_{nlt}\int K_t(\mu-\nu)\tau_l(\nu)d\nu,
    \nonumber\\
\tau_n+\tau_n^h=&&\sum_{lt}B_{nlt}\int K_t(\nu-\mu)\omega_l(\mu)d\mu
 -\sum_{lt}A_{nlt}\int K_t(\nu-\nu^\prime)\tau_l(\nu^\prime)d\nu'.
\label{Integralequations}
\end{eqnarray}
where $K_n(x)=\pi^{-1}n\eta/(n^2\eta^2+x^2)$,
and the integration limits are defined by
\begin{eqnarray}
\frac{N}{L}=&&\int_{-Q_0}^{Q_0}\rho(k) dk,\nonumber\\
\frac{M}{L}=&&\sum_nn\int_{-B_1}^{B_1}\sigma_n(\lambda) d\lambda, \nonumber\\
\frac{M'}{L}=&&\sum_nn \int_{-B_2}^{B_2}\omega_n(\mu) d\mu,\nonumber\\
\frac{M''}{L}=&&\sum_nn\int_{-B_3}^{B_3}\tau_n(\nu) d\nu.
\label{NMMM}
\end{eqnarray}
Since $|1\rangle=|\tup\rangle$, $|2\rangle=|\tdown\rangle$,
$|3\rangle=|\bup\rangle$ and $|4\rangle=|\bdown\rangle$,
the $z$-components of total spin and orbital are given by,
\begin{eqnarray}
\frac{S_z}{L}&=&\frac{1}{2}\int\rho dk+\sum_nn\int\omega_n d\mu
 -\sum_nn\int\sigma_n d\lambda
  -\sum_nn\int\tau_n d\nu,
    \nonumber\\
\frac{T_z}{L}&=&\frac{1}{2}\int \rho dk-\sum_nn\int\omega_n d\mu.
\label{STZ}
\end{eqnarray}
The energy is given by
\begin{equation}
\frac{E}{L}=-2t\int \cos k\rho(k)dk,
\end{equation}
and the magnetization by
\begin{equation}
M_z=g_sS_z+g_tT_z,
\label{EQ:MAG}
\end{equation}
where $g_s,g_t$ are Land\'e $g$ factors that we know $g_s=2, g_t=1$,
which was ignored in previous literature because it was not
able to be related to spin-orbital model.

\section{Thermal equilibrium}
\label{sec:3}

Yang and Yang\cite{Yang} first introduced a definition of entropy 
in terms of distribution of roots and holes
for Bethe-ansatz solvable systems.
This method was recently carefully compared with the transfer matrix
approach, the results of both approach are shown in agreement with each
other \cite{DEGKKK}.
In order to obtain thermal equilibrium at finite temperature, 
we should maximize the free energy $\Omega=E+E_J-T{\cal S}-{\cal A}N$ 
where $E_J=-\mu_0 HM_z$ is the Zeemann energy due to the 
external magnetic field and ${\cal A}$ is the chemical potential.
The entropy is given by
\begin{eqnarray}
\frac{\cal S}{L}=&&\int[(\rho+\rho^h)\ln(\rho+\rho^h)
 -\rho\ln\rho-\rho^h\ln\rho^h]dk 
  \nonumber\\
 &&+\sum_n\int[(\sigma_n+\sigma_n^h)\ln(\sigma_n+\sigma_n^h)
 -\sigma_n\ln\sigma_n
  -\sigma_n^h\ln\sigma_n^h]d\lambda 
    \nonumber\\
 &&+\sum_n\int[(\omega_n+\omega_n^h)\ln(\omega_n+\omega_n^h)
  -\omega_n\ln\omega_n-\omega_n^h\ln\omega_n^h]d\mu \nonumber\\
    &&+\sum_n\int[(\tau_n+\tau_n^h)\ln(\tau_n+\tau_n^h)-\tau_n\ln\tau_n
    -\tau_m^h\ln\tau_n^h]d\nu.
\label{Entropy}
\end{eqnarray}
We obtain the Helmholtz free energy from the condition $\delta \Omega=0$ that
\begin{equation}
F=-\frac{TL}{2\pi}\int\ln[1+e^{-\epsilon}]dk,
\label{FreeEnergy}
\end{equation}
where $\epsilon$ should be solved from the following equations,
\begin{eqnarray}
\epsilon=&&-\frac{1}{T}\Bigl[2t\cos k+{\cal A}+\frac{3}{2}\mu_0 H \Bigr]
 -\sum_n\int K_n(\lambda-\sin k)\ln[1+e^{-\theta_n(\lambda)}]
  d\lambda,
   \nonumber\\
\theta_n=&&2\mu_0 n H/T-\int\cos kK_n(\sin k-\lambda)
 \ln[1+e^{-\epsilon(k)}]dk  \nonumber \\
  &&+\sum_{lt}A_{nlt}\int K_t(\lambda-\lambda')
  \ln[1+e^{-\theta_l(\lambda')}]d\lambda'
 -\sum_{lt}B_{nlt}\int K_t(\mu-\lambda)\ln[1+e^{-\zeta_l(\mu)}]d\mu, 
  \nonumber\\
\zeta_n=&&-\mu_0nH/T-\sum_{lt}B_{nlt}\int K_t(\lambda-\mu)
 \ln[1+e^{-\theta_l(\lambda)}]d\lambda 
  \nonumber\\
&&+\sum_{lt}A_{nlt}\int K_t(\mu-\mu')\ln[1+e^{-\zeta_l(\mu')}]
  d\nu'-\sum_{lt}B_{nlt}\int K_t(\nu-\mu)\ln[1+e^{-\xi_l(\nu)}]d\nu, 
  \nonumber\\
\xi_n=&&2\mu_0nH/T-\sum_{lt}B_{nlt}\int K_t(\mu-\nu)
 \ln[1+e^{-\zeta_l(\mu)}]d\mu
  \nonumber\\
&&+\sum_{lt}A_{nlt}\int K_t(\nu-\nu')
 \ln[1+e^{-\xi_l(\nu')}]d\nu'.
\label{ThermalEquation}
\end{eqnarray}
In getting the above expressions, we have introduced notations:
\begin{equation}
e^\epsilon=\frac{\rho^h}{\rho}, \;\; 
e^{\theta_n}=\frac{\sigma_n^h}{\sigma_n}, \;\;
e^{\zeta_n}=\frac{\omega_n^h}{\omega_n},\;\;
e^{\xi_n}=\frac{\tau_n^h}{\tau_n}.
\label{DHD}
\end{equation}

\section{The case at low temperature}
\label{sec:4}

At very low temperature, the density of roots undergo a slight modification 
from that of the ground state, and the right hand side of Eq. (\ref{DHD}) are
approximately zero below Fermi surface. Then we have
\begin{eqnarray}
\epsilon=&&-\frac{1}{T}\Bigl[2t\cos k+{\cal A}+\frac{3}{2}\mu_0 H\Bigr]+
\int_{-B_1}^{B_1} K_1(\lambda-\sin k)\theta(\lambda)d\lambda,\nonumber \\
\nonumber
\theta=&&2\mu_0H/T+\int_{-Q_0}^{Q_0} \cos kK_1(\sin k-\lambda)\epsilon(k)dk
 -\int_{-B_1}^{B_1} K_2(\lambda-\lambda')\theta(\lambda')
  d\lambda'+\int_{-B_2}^{B_2} K_1(\mu-\lambda)\zeta(\mu)d\mu, \\
\nonumber
\zeta=&&-\mu_0H/T+\int_{-B_1}^{B_1} 
  K_1(\lambda-\mu)\theta(\lambda)d\lambda-\int_{-B_2}^{B_2} K_2(\mu-\mu')
  \zeta(\mu')d\mu'+\int_{-B_3}^{B_3} K_1(\nu-\mu)\xi(\nu)d\nu,\\
\xi=&&2\mu_0H/T+\int_{B_2}^{B_2} K_1(\mu-\nu)\zeta(\mu)d\mu-
    \int_{-B_3}^{B_3} K_2(\nu-\nu')
  \xi(\nu')d\nu'.
\label{TES}
\end{eqnarray}
It is plausible to take $B_1, B_2, B_3=\infty$ in our present case.
By making a Fourier transform of the second through
fourth equation of Eq. (\ref{TES}), we have
\begin{eqnarray}
\tilde{\theta}&&=2\tilde{h}+\frac{1}{2\pi}\int_{-Q_0}^{Q_0}\cos k
  e^{-\eta|w|-iw\sin k}\epsilon(k)dk
   -e^{-2\eta|w|}\tilde{\theta}+e^{-\eta|w|}\tilde{\zeta},\\
 \nonumber
\tilde{\zeta}&&=-\tilde{h}+e^{-\eta|w|}\tilde{\theta}
  -e^{-2\eta|w|}\tilde{\zeta}+e^{-\eta|w|}\tilde{\xi},\\
\nonumber
\tilde{\xi}&&=2\tilde{h}+e^{-\eta|w|}\tilde{\zeta}
  -e^{-2\eta|w|}\tilde{\xi}
\end{eqnarray}
where $\tilde{h}=\mu_0H\delta(w)/T$. It is not difficult to
derive an integral equation for $\epsilon$:
\begin{equation}
\epsilon=-[2t\cos k+{\cal A}]/T+\frac{1}{\eta}
  \int_{-k_F}^{k_F}
  \cos k^\prime \epsilon(k^\prime)
   R_3\Bigl(\frac{\sin k-\sin k^\prime}{\eta}\Bigr)dk
\label{LTE}
\end{equation}
where $k_F$ is the Fermi momentum and
\begin{equation}
R_n(x)=\frac{1}{2\pi}\int_{-\infty}^\infty \frac{\sinh (nw)}{\sinh (4w)}
      e^{iwx-|w|}dw.
\label{RF}
\end{equation}

In the absence of external field ($H=0$), Eq.(\ref{LTE}) is rigorous. When
the external field is weak, it is also plausible because
the changes on the integral interval of spin and orbit rapidity is of
order of $1/L$ in the thermodynamic limit. Moreover, even if we consider
the contribution of the densities outside the Fermi surface, it is of
order $e^{-1/T}$. Clearly the presence of external
field has no explicit effect on the ratio of density of charge roots and holes
at low temperature. We can also infer from Eq.(\ref{LTE}) and Eq. (\ref{EQ:MAG}) there will
be two critical fields $H_c^1, H_c^2$
which can cause phase transition on the magnetization of the system.

Once $\epsilon$ is solved, the free energy at low temperature can be
evaluated by Eq.(\ref{FreeEnergy}). Though an explicit expression
can not be obtained from Eq. (\ref{LTE}) in general case,
it is easier for numerical calculation.
Define the density of roots at low temperature as $\rho=\rho_0+\rho_1$,
where $\rho_1$ is a slight deviation from the density of 
the ground state $\rho_0$ \cite{LiGYE00},
we can obtain
\begin{equation}
\rho=\rho_0(1-e^\epsilon)-\frac{1}{\pi}\int\frac
  {\cos k\eta\sigma_0e^\theta d\lambda}{\eta^2+(\sin k-\lambda)^2}.
\end{equation}

Substituting  $\epsilon(k)$ of Eq. (\ref{LTE}) into Eq. (\ref{FreeEnergy}),
we obtain
\begin{equation}
F=\frac{TL}{2\pi}\int\epsilon dk-\frac{TL}{2\pi}\int e^\epsilon dk.
\end{equation}
Clearly when $T\rightarrow 0$ the free energy is just the ground state energy
except for the additional
chemical potential term $-{\cal A}L$ and the $\epsilon(k)$ just play a role
of so called dressed energy.

{\bf Strong coupling limit}

In strong coupling limit $\eta\rightarrow \infty$, we have
\begin{equation}
\epsilon=-[2t\cos k+{\cal A}]/T,
\label{TTT}
\end{equation}
Since we consider the problem at low concentration, the Fermi surface
is much less than $\pi$. Then we can rewrite
Eq. (\ref{TTT}) as
\begin{equation}
\epsilon=[tk^2-\kappa]/T,
\end{equation}
where $\kappa=2t+{\cal A}$. And the distribution of the roots becomes

\begin{equation}
\rho=\frac{1}{2\pi}\frac{1}{1+e^{[tk^2-\kappa]/T}},
\end{equation}
with condition $\int_{-k_F}^{k_F}\rho(k) dk=D$, and when $T=0$,
it gives $k_F=\sqrt{\kappa_0 /t}$. We obtain
\begin{equation}
\kappa=\kappa_0\Bigl[1-\frac{\pi^2 T^2}{24\kappa_0^2}\Bigr]^{-2},
\end{equation}
where $\kappa_0=t\pi^2D^2$ is the value of $\kappa$ at zero temperature.
Obviously the system's energy becomes
\begin{eqnarray}
\frac{E}{L}\doteq \frac{\kappa_0^{3/2}}{3\pi\sqrt{t}}
  \Bigl[1+\frac{\pi^2 T^2}{4\kappa_0^2}\Bigr].
\end{eqnarray}

The specific heat and the entropy of the system are calculated as
\begin{equation}
C_V={\cal S}=\frac{T}{6tD},
\end{equation}
which is Fermi-liquid like.

{\bf Weak coupling limit}

In weak coupling $\eta\rightarrow 0$, Eq.(\ref{LTE}) can be
solved exactly. In mathematics we have
$$
\lim_{\eta\rightarrow 0}\frac{\sinh (3\eta)}{\sinh (4\eta)}e^{-\eta}
=\frac{3}{4},
$$
Then the Eq. (\ref{RF}) is proportional to $\delta$-function. The
$\epsilon(k)$ has the form
\begin{equation}
\epsilon=-[8t\cos k+4{\cal A}]/T.
\end{equation}
So we can repeat the steps in the section of strong coupling limit.
The specific heat and the entropy become
\begin{equation}
C_V={\cal S}=\frac{T}{96tD}
\end{equation}

Clearly it is still Fermi-liquid like. We can infer at low temperature
the specific heat of the system is Fermi-liquid like regardless the
value of coupling constant. The difference
is the coefficient.

{\bf Magnetic properties}

In the presence of external field, the densities of spin and
orbit rapidities will change. We rewrite them as 
\begin{eqnarray}
\sigma=\sigma_0(1-e^\theta),\nonumber  \\
\omega=\omega_0(1-e^\zeta),\nonumber \\
\tau=\tau_0(1-e^\xi).
\end{eqnarray}
where $\theta, \zeta, \xi$ are all negative and proportional to $H/T$.
Then the magnetization of the system has the form
\begin{equation}
M_z/L=\frac{3}{2}\int \rho dk+\int \omega d\mu-2\int \sigma d\lambda
-2\int \tau d\nu.
\end{equation}

Concerning $M=3N/4, M'=N/2, M''=N/4$ at the ground state, we obtain
\begin{equation}
M_z/L=2\int \sigma_0 e^\theta d\lambda
  -\int \omega_0 e^\zeta d\mu+2\int \tau_0e^\xi d\nu.
\end{equation}
and the susceptibility in the weak external field is
\begin{equation}
\chi=\frac{3\mu_0}{T}\int \sigma_0e^\theta d\lambda
  -\frac{\mu_0}{T}\int \omega_0 e^\zeta d\mu
    +\frac{3\mu_0}{T}\int \tau_0 e^\xi d\nu.
\label{CHI}
\end{equation}

However, if the external field is strong enough so that there
are no spin and orbit flipped down. That is $M=M'=M''=0$ 
and the densities of $\lambda, \mu, \nu$ are all zero.
Then we have
\begin{equation}
\rho+\rho^h=\rho(1+e^\epsilon)=\frac{1}{2\pi},
\end{equation}
where $\epsilon$ is given by
\begin{eqnarray}
\epsilon=-\frac{1}{T}\Bigl[2t\cos k+{\cal A}+\frac{3}{2}\mu_0H\Bigr],\nonumber \\
2\mu_0H_c/T=-\int \cos k K_1(\sin k-\lambda)\epsilon dk.
\end{eqnarray}
The $H_c$ is the maximal value of $\{H_c^1, H_c^2\}$ mentioned
above. At large coupling constant the value of $H_c$ scales like
$1/\eta$.
Clearly, in the limit $\eta\rightarrow \infty$, a small value of
$H$ can flip all spin and orbit to upward. The density of charge roots
becomes
\begin{equation}
\rho=\frac{1}{2\pi}\frac{1}{1+e^{-[2t\cos k+2t]/T}}
\label{DDD}
\end{equation}
which takes the value of $1/2\pi$ at the ground state.
The density of charge roots in strong coupling limit takes the same form
as that in sufficient strong external field.

\section{Summary and acknowledgments}

In the above, we discussed one-dimensional SU(4) Hubbard-like Model in 
the thermodynamic limit, which is reliable at low temperature and
low concentration.
We studied the property of the thermal equilibrium
and obtained the equilibrium equations. 
In the low temperature limit,
the T-dependent density of roots and free energy, 
specific heat both in strong and weak coupling limit, as well
as magnetic properties of the system are calculated. 

This work is supported by NSFC No.1-9975040, trans-century project
 and EYF of China Education Ministry. 
Y.Q.Li acknowledges the support of AvH-stiftung.
S.J.Gu acknowledges interesting discussions with J.H.Dai and D.Yang.

\end{document}